\documentclass[reprint, 10pt, floatfix,aps,prd,twocolumn,a4paper,superscriptaddress,nofootinbib,notitlepage]{revtex4-1}
\usepackage[utf8]{inputenc}
\usepackage[T1]{fontenc} 
\usepackage{paralist}
\usepackage{hyperref}
\usepackage[english]{babel}
\usepackage{amsopn,amsthm,dsfont,amssymb}
\usepackage{mathrsfs}
\usepackage{cancel} 
\usepackage{comment} 
\usepackage{color} 
\usepackage{soul} 
\usepackage[normalem]{ulem}
\usepackage{cancel} 
\usepackage{hyperref} 
\hypersetup{
    colorlinks=true,
    linkcolor=blue,
    filecolor=magenta,      
    urlcolor=cyan,
}
\usepackage{mathrsfs}
\usepackage{amsmath}

\usepackage{mathtools}

\usepackage[utf8]{inputenc}
\usepackage[T1]{fontenc} 
\usepackage[english]{babel}
\usepackage{amsopn,amsthm,dsfont,amssymb}
\usepackage{mathrsfs}
\usepackage{cancel} 
\usepackage{comment} 
\usepackage{color} 
\usepackage{soul} 
\usepackage[normalem]{ulem}
\usepackage{siunitx}

\usepackage{tikz}
\usetikzlibrary{positioning,calc,arrows,decorations.pathmorphing,shapes}
\usepackage{adjustbox}
\usepackage[caption=false]{subfig}

\AtBeginDocument{%
    \newwrite\bibnotes
    \def\bibnotesext{Notes.bib}
    \immediate\openout\bibnotes=\jobname\bibnotesext
    \immediate\write\bibnotes{@CONTROL{REVTEX41Control}}
    \immediate\write\bibnotes{@CONTROL{%
    apsrev41Control,author="08",editor="1",pages="1",title="0",year="1"}}
     \if@filesw
     \immediate\write\@auxout{\string\citation{apsrev41Control}}%
    \fi
}%

\DeclareMathOperator{\Tr}{Tr}
\newcommand{\ket}[1]{|#1\rangle}
\newcommand{\bra}[1]{\langle #1|}
\newcommand{\braket}[2]{\langle #1 | #2 \rangle}

\begin{document}
\title{Mass-energy equivalence in gravitationally bound quantum states of the neutron}
\author{Germain Tobar}
\affiliation{Centre for Engineered Quantum Systems, School of Mathematics and Physics,\\The University of Queensland, St.\ Lucia, QLD 4072, Australia}
\affiliation{Department of Applied Mathematics and Theoretical Physics,
Centre for Mathematical Sciences,
Wilberforce Road,
Cambridge, CB3 OWA, UK}
\author{Simon Haine}
\affiliation{Department of Quantum Science, Research School of Physics,\\The Australian National University, ACT 0200, Australia}
\author{Fabio Costa}
\affiliation{Centre for Engineered Quantum Systems, School of Mathematics and Physics,\\The University of Queensland, St.\ Lucia, QLD 4072, Australia}
\author{Magdalena Zych}
\affiliation{Centre for Engineered Quantum Systems, School of Mathematics and Physics,\\The University of Queensland, St.\ Lucia, QLD 4072, Australia}
\begin{abstract}
Gravitationally bound neutrons have become an important tool in the experimental searches for new physics, such as modifications to Newton's force or candidates for dark matter particles. Here we include the relativistic effects of mass-energy equivalence into the model of gravitationally bound neutrons. Specifically, we investigate a correction in a gravitationally bound neutron's Hamiltonian due to the presence of an external magnetic field. We show that the neutron's additional weight due to mass-energy equivalence will cause a small shift in the neutron's eigenenergies and eigenstates, and examine how this relativistic correction would affect experiments with trapped neutrons. We further consider the ultimate precision in estimating the relativistic correction to the precession frequency and find that, at short times, a joint measurement of both the spin and motional degrees of freedom provides a metrological enhancement as compared to a measurement of the spin alone.
\end{abstract}
\maketitle

\section{Introduction}

The theories of general relativity (GR) and quantum mechanics are the two pillars of modern physics. However, the unification of these two theories remains an unsolved problem. Due to their differences, it is expected that the exploration of the experimentally untested regime where both theories apply may lead to the discovery of new physics \cite{book, Kieferbook, HowlFuentes:2018GravLab}. {There is therefore strong interest in physical systems suitable for high precision experiments in which joint quantum and gravitational phenomena could be tested. In this context cold neutrons offer a particularly interesting perspective: Historically neutrons allowed the first experimental demonstration of a coherent phase shift due to gravitational potential on a spatial superposition of a particle~\cite{cow}, and they are the only system for which quantisation of bound states in the gravitational potential well has been experimentally demonstrated~\cite{paper1,  Nesvizhevsky05, natureexperiment, Kamiya, Ichi}.}
%
{These advances} have led to the application {of neutrons} in tests of fundamental physics such as the search for dark matter particles {e.g.~}axions \cite{axion1, darkmatter2} {or chameleon fields~\cite{PhysRevD.93.062001, Brax:2014testingChameleons, Burrage:2018LivRevChameleon}, and short-distance modifications of the Newtonian gravitational force~\cite{ABELE:2009QuBounce}. They are also promising for future tests of Lorentz violations~\cite{Abele:2019SME}}. 
All tests involving gravitationally bound neutrons were so far conducted in a regime where {the framework of Newtonian gravity was sufficient to explain the results~\cite{Nesvizhevskyreview}}. As tests the fundamental physics demand higher levels of precision, {general relativistic effects on neutrons in these tests will become relevant}.  


In this work, we investigate the effects of mass-energy equivalence in gravitationally bound quantum states of the neutron. 
Mass-energy equivalence becomes relevant as we consider the internal energy of the neutron in an additional external magnetic field. {The new aspect of our considerations is that the resulting internal energy contributions to the mass are quantised, as they are given by the neutron spin projection on the magnetic field}. We derive the resulting relativistic corrections to gravitationally bound neutron states and to their eigenenergies. We discuss the feasibility of the use of {interferometric techniques} to detect these relativistic corrections. We show that if the experiment were to be performed with a sufficiently large number of neutrons, the sensitivity required to detect the quantum effects of the mass-energy equivalence can in principle be achieved {with present-day neutron sources}. Finally, we formulate the task of detecting the effects of mass-energy equivalence as a quantum parameter estimation problem and show that with an appropriate experimental procedure, the effects of mass-energy equivalence can in principle have a much more significant impact in high precision experiments than they would in a simple interferometric measurement of phase.

The study of mass-energy equivalence in gravitationally bound neutrons is further motivated by their potential for a novel test of the Einstein equivalence principle (EEP) in a quantum context. The gravitationally bound neutron states simultaneous dependence on the gravitational and inertial mass-energies makes them a suitable candidate for testing the quantum formulation of the EEP~\cite{QEP,Orlando2016,Anastopoulos2018}, {where accessing the quantised mass-energies of test particles is crucial}.

\section{Gravitationally bound neutrons} \label{sec2}
This section summarises {the relevant aspects of gravitationally bound neutrons in the non-relativistic limit following} Ref.~\cite{Nes}.

\begin{figure}[h]
\begin{center}
\includegraphics[scale  = 0.22]{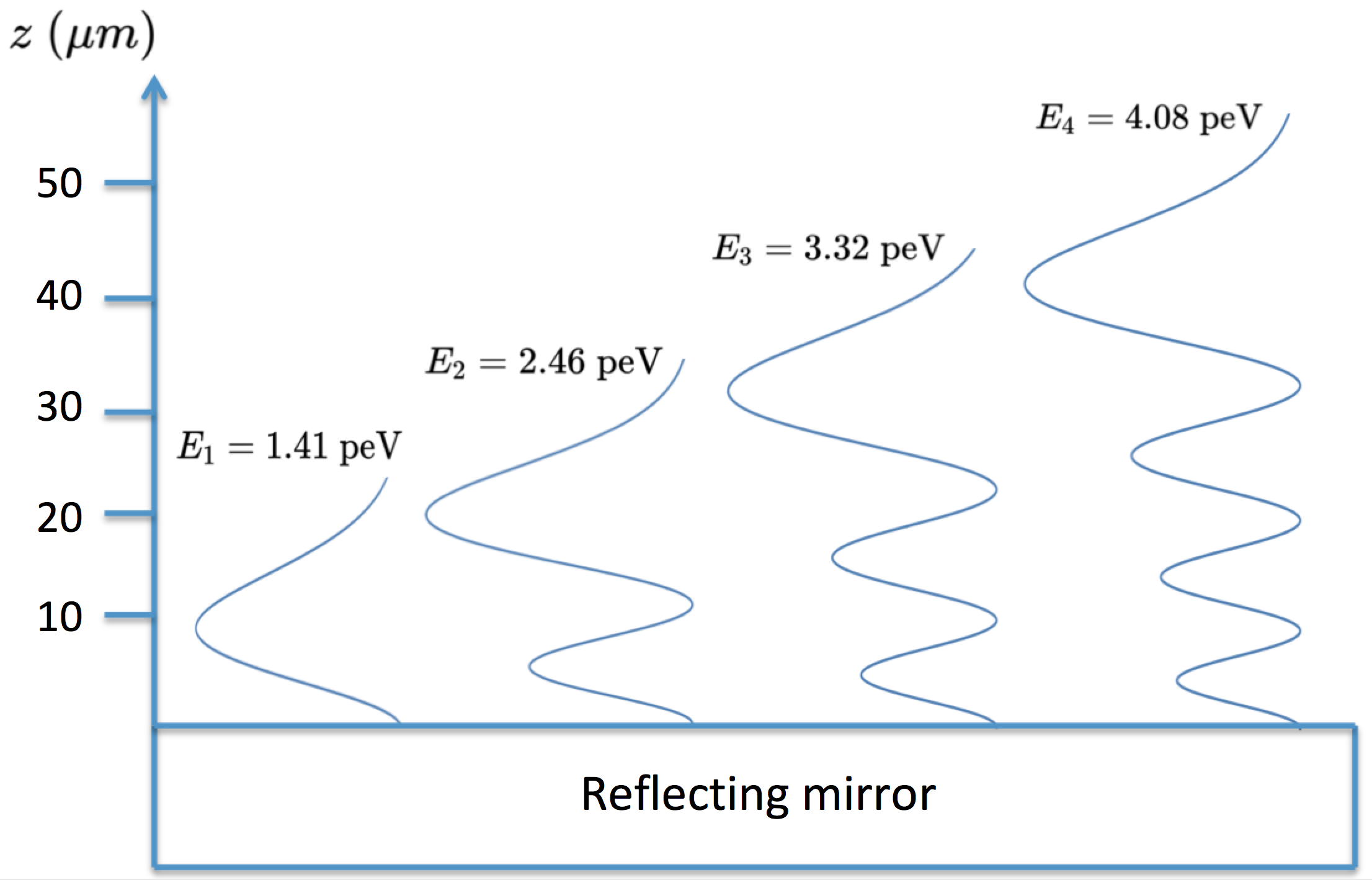}
\caption{\label{airyfn} {Energy eigenstates  $\ket{\psi_n}$ 
of a neutron in a potential due to Earth's gravitational field and a reflecting mirror. The wave functions are displayed in position basis, where the coordinate $z$ represents the vertical height above the  mirror. $E_n$ denotes the energy of the $n^\mathrm{th}$ eigenstate.}}
\end{center}
\end{figure}

Consider a neutron in a gravitational potential in the $z$-direction and above a reflecting mirror, where the mirror acts as an infinite potential barrier. The resulting potential is
\begin{equation} \label{potential}
   V( z) = \left \{
\begin{array}{ll}
      \infty & z < 0 \\
      g z & z\geq 0 \\
\end{array} 
\right \}.
\end{equation}
This potential generates quantum states of the neutron which are described by the Hamiltonian
\begin{equation} \label{hamiltonian}
    \hat{H_1} = \frac{\hat{p}^2}{2m} + mV(\hat{z}),
\end{equation}
where $m$ is mass of the neutron. The corresponding stationary states $\ket{\psi_n}$ are
\begin{equation} \label{airy}
    \braket{z}{\psi_n} = A_n\phi\left(\frac{z}{\lambda} + \gamma_n\right),
\end{equation}
where $\phi(x)$ is an Airy function, $A_n$ is a normalization constant, $\lambda := \sqrt[3]{\hbar^2/2m^2g}$ is a length scale for the quantum states and $\gamma_n $ are the zeros of the Airy function (solutions to $\phi(x) = 0$), see Fig.~\ref{airyfn}. The corresponding eigenenergies $E_n$ are solutions of the following equation:
\begin{equation} \label{energies}
    \phi\left(-\frac{\sqrt[3]{2}}{\sqrt[3]{mg^2\hbar^2}}E_n\right) = 0.
\end{equation}
The four lowest energy levels are $1.41$, $2.46$, $3.32$ and $4.08$~{peV} for the gravitational acceleration at the surface of the Earth.

These discrete energy levels demonstrate how a Newtonian gravitational potential in combination with a reflecting mirror generates quantised states of the neutron.  In {the next section} we examine relativistic corrections to these states as a probe for novel effects at the interplay of general relativity and quantum mechanics. 

\section{Mass-energy equivalence in quantum systems}
In order to consider the effects of mass-energy equivalence in neutrons, we must firstly examine how the quantisation of mass-energy will affect the neutron's Hamiltonian. A previously developed formulation~\cite{Visibility, zych2015PhD, Pikovski2017, QEP, zych2018gravitational,GiuliniSchwartz2019,RouraPRX2020} describes {dynamics of particles with quantized internal energy} in terms of a mass-energy operator
\begin{equation}\label{MEQ}
    \hat{M} = m\hat{I} + \frac{\hat{H}_{\textrm{int}}}{c^2}.
\end{equation}
Here, $\hat{H}_{int}$ is the Hamiltonian due to internal dynamics, and $m$ is the particle's uncorrected rest mass. If we consider the presence of a constant external magnetic field aligned along the $z$ axis then the interaction of the neutron with the magnetic field will become the neutron's internal Hamiltonian 

\begin{equation} \label{interaction}
    \hat{H}_{\textrm{int}} = \frac{\hbar \omega_0}{2}\hat{\sigma}_z,
\end{equation}
where, $\omega_0$ is the neutron's Larmor frequency. If we consider the neutron to simultaneously be in the presence of the gravitational potential described by eq.~\eqref{potential}, then {at low centre of mass energies} the neutron's Hamiltonian {reads}
\begin{equation} \label{general}
    \hat{H} = \hat{M}c^2 + \frac{\hat{p}^2}{2\hat{M}} + \hat{M}V(\hat{z}),
\end{equation}
with the mass-energy $\hat M$ given by eqs~\eqref{MEQ} and \eqref{interaction}.  The spin degree of freedom (DOF) is of course not a Lorentz scalar like the rest mass of the particle. However, in the considered regime of slow Centre of Mass (COM) motion, and to lowest order  relativistic corrections coming from internal DOFs, the energy associated with the spin effectively contributes to the mass of the particle as per the mass-energy equivalence~\cite{Brown1986, MORISHIMA2004, bushev2016single}. 

We denote the {eigenstates of} the Hamiltonian {eq.~}\eqref{general} to be $\ket{\Psi_{n,s}} = \ket{\psi_{n,s}}\ket{s}$, where $\ket{\psi_{n,s}}$ describes the neutron's {position wave function} and $\ket{s}$ describes the neutron's spin state. Note, that we can express the eigenstates as a product of the gravitational and the spin states as the mass-energy operator commutes with the external DOF. The above notation recognises that the quantisation of mass-energy in the Hamiltonian will result in gravitational eigenstates which depend on the neutron's spin state, and therefore are slightly different from the states described in eq.~\eqref{airy}. 
We  denote the corresponding eigenenergies to be $E_{n,s}^{\textrm{total}}$. For example, $\ket{\Psi_{1,\uparrow}} = \ket{\psi_{1,\uparrow}}\ket{\uparrow}$ describes a spin up neutron in the ground gravitational state with eigenenergy $E_{1,\uparrow}^{\textrm{total}}$. 

The eigenenergies $E^{\textrm{total}}_{n,s}$ of the Hamiltonian \eqref{general} {include} energy due to the interaction of the neutron{'s spin} with the external magnetic field $\pm \frac{\hbar \omega_0}{2}$, as well as the gravitational energy, which we denote $E_{n,s}$. This gravitational energy can be found by {appropriately} shifting the mass of the neutron in equation \eqref{energies} {by including} mass-energy equivalence:
\begin{equation} \label{energies2}
    \begin{split}
    \phi\left(-\frac{\sqrt[3]{2}}{\sqrt[3]{(m + \frac{\hbar \omega_0}{2c^2})g^2\hbar^2}}E_{n,\uparrow}\right) = 0,\\
    \phi\left(-\frac{\sqrt[3]{2}}{\sqrt[3]{(m - \frac{\hbar \omega_0}{2c^2})g^2\hbar^2}}E_{n,\downarrow}\right) = 0.    
    \end{split}
\end{equation}
As shown in Appendix~\ref{app2}, if we apply a binomial approximation for {$\hbar\omega_0\ll mc^2$} we obtain a simple expression for the relativistic correction to the neutron's {non-relativistic} eigenenergies $E_n$:
\begin{equation} \label{energy_corrected}
E_{n, s} = E_n\left(1 \pm \frac{\delta}{3}\right),
\end{equation}
where 
\begin{equation}
	\delta =  \frac{\hbar \omega_0}{2mc^2}
\end{equation}
is a dimensionless {relativistic correction} and where the sign $\pm$ depends on spin  state, {i.e.~$ s = \uparrow(\downarrow)$ corresponds to $+(-)$}. The same result can be derived from  perturbation theory to first order in $\delta$, see Appendix~\ref{airy_a}.  The total energy of a neutron with a Hamiltonian in eq.~\eqref{general} is the sum of the neutron's spin energy and gravitational energy,
\begin{equation} \label{energy_total}
\begin{split}
        E_{n, s}^{\textrm{total}} &= E_{n}\left(1 \pm \frac{\delta}{3} \right) \pm \frac{\hbar \omega_0}{2},\\
          &= E_{n} \pm \frac{\hbar \omega_{0}}{2}\left(1+\frac{E_{n}}{3 m c^{2}}\right).
\end{split}
\end{equation}
Here, it is clear that the relativistic correction can either be viewed as a spin state dependent shift to the gravitational energy levels or a gravitational energy level dependent shift to the spin energy levels. Equation \eqref{energy_total} describes the eigenenergies of a {slow} neutron (whose centre of mass in the laboratory reference frame is non-relativistic) in a gravitational potential and a constant magnetic field.


\subsection{Quantitative analysis of the relativistic corrections} \label{sec4}
The relativistic effect in equation \eqref{energies2} is given by the additional mass $\hbar\omega_0/2c^2$ due to the spin energy in the external magnetic field, {where the Larmor frequency $\omega_0$ grows with the magnetic field strength}. The strongest continuous magnetic field produced in a laboratory {to date} is $45$~T~\cite{highmagfield}, the strongest pulsed magnetic field produced in a laboratory is $1200$~T \cite{highermagfield}. {For comparison,}  the magnetic field of a neutron star is typically on the order of $10^7$~T~\cite{neutronstar} -- these stars provide the most extreme magnetic fields so far observed in nature. These magnetic field strengths still {satisfy} $\hbar\omega_0/c^2 \ll  m$. In Table~\ref{exp_results} we calculate the difference between the eigenenergies with, eq~\eqref{energy_corrected}, and without the relativistic corrections, {i.e.~}$E_{n,\uparrow} - E_{n} = \frac{\delta}{3}E_n$, for various field strengths. For example, a constant magnetic field of $45$~T produces a relativistic correction factor given by $\delta \approx 2.88 \times 10^{-15}$. This single parameter characterizes the relativistic correction to each gravitational energy level.

\begin{table}[ht] 
\begin{center}
\begin{tabular}{|m{1cm}|m{1.4cm}|m{1.4cm}|m{1.4cm}|m{1.4cm}|} 
\hline
$B$ (Tesla)& $E_{1, \uparrow} - E_{1}$ (peV)& $E_{2, \uparrow} - E_{2}$ (peV)& $E_{3, \uparrow} - E_{3}$ (peV)& $E_{4, \uparrow} - E_{4}$ (peV)\\
\hline
 45 & $1.36\mathrm{e}{-15}$  &  $2.37\mathrm{e}{-15}$ & $3.20\mathrm{e}{-15}$ & $3.94\mathrm{e}{-15}$\\
\hline
1200 & $3.62\mathrm{e}{-14}$  & $6.32\mathrm{e}{-14}$ & $8.54\mathrm{e}{-14}$ & $1.05\mathrm{e}{-13}$\\
\hline
$10^{7}$ & $3.01\mathrm{e}{-10}$&$5.27\mathrm{e}{-10}$& $7.12\mathrm{e}{-10}$& $8.75\mathrm{e}{-10}$\\
\hline
\end{tabular}
\caption{\label{exp_results} The energy difference $E_{1, \uparrow} - E_{1} = \frac{\delta E_n}{3}$ between a gravitationally bound neutron's eigenenergies with and without the contribution of internal energy to the neutron mass. The energy differences are calculated for magnetic fields of $45$, $1200$ and $10^7$ Tesla. The calculations for the first our gravitational energy levels are displayed. }
\end{center} 
\end{table}

The eigenenergies of the neutron are on the order of peV as calculated in section \ref{sec2}. In order to distinguish the contribution of internal energy to {the neutron} mass {in} an external magnetic field of $45$~T, we will need to be able to determine the neutron's energies with a resolution on the order of $10^{-16}$ peV. If the experiment were to be conducted with a magnetic field strength of $1200$ T, the eigenenergies would need to be resolved at the level of $10^{-14}$ peV. Even with a magnetic field strength comparable to that of a neutron star, the shifts in the neutron's eigenenergies are very small, on the order of $10^{-10}$ peV. The size of these relativistic corrections confirms that in present-day experiments these effects are negligible. However, as precision measurement improves, it may become feasible to measure the neutron's energy to this precision. The potential for these effects to influence future high-precision experiments, as well as provide a route to a novel test of the quantum formulation of the EEP~\cite{Orlando2016, QEP}, makes it important to consider how these relativistic effects can be experimentally measured.

\section{Experimental procedure} \label{sec5}
One possible avenue to detect the relativistic corrections to the neutron's eigenenergies is through an interferometric measurement.

\begin{figure}[ht!]
\begin{center}
\includegraphics[scale  = 0.37]{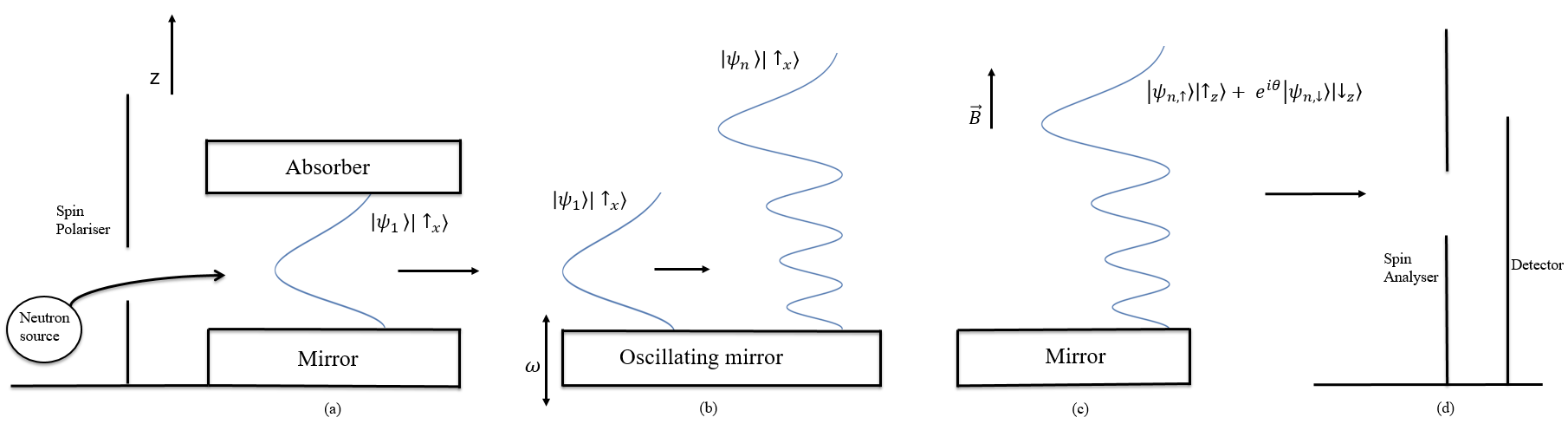}
\caption{\label{experiment} {Experimental setup using spin interferometry to detect the perturbation to the gravitationally bound neutron's eigenenergy resulting from the mass-energy equivalence.} (a) In the first region,  spin-polarised neutrons enter a \textit{ground state selector} in which we use a polished mirror in combination with a rough absorber/scatterer to select neutrons in the gravitationally bound ground state. To achieve this, the absorber is placed at a height of 20 $\mu$m, such that neutrons that are not in the ground state are scattered out of the setup. (b) In the second region, we use the process described in Ref.~\cite{mechoscillation} which involves the application of sinusoidal mechanical (Rabi) oscillations to the remaining ground state neutrons at a frequency resonant with the transition $\ket{\psi_{1}}\ket{\uparrow_x} \rightarrow \ket{\psi_{n}}\ket{\uparrow_x}$. The application of a $\pi$ pulse in this way prepares the neutrons in the excited state $\ket{\psi_{n}}\ket{\uparrow_x}$. (c) In the third region this state evolves in the presence of a constant magnetic field along the z-axis which causes the spin to precess but without exerting a force and thus without changing the state of the position degree of freedom. (d) {Finally, the phase $\theta$ acquired during the previous stage is measured (from the population difference between spin projections in the $x$ basis).} }
\end{center}
\end{figure}

Our proposed experiment consists of four stages, see Fig.~\ref{experiment}.  
The first stage selects spin polarised neutrons which are in the ground gravitational energy level out of the initial ensemble coming from a neutron source. If we select neutrons spin-polarised in the positive $x$ direction, the initial state of the system is $|\Psi_0\rangle = |\psi_1\rangle |\uparrow_x\rangle$. In the second stage, 
{sinusoidal mechanical oscillations at the frequency $\omega_{1,n}$,  resonant with the transition $\ket{\psi_1}~\ket{\uparrow_x} \rightarrow \ket{\psi_n}\ket{\uparrow_x}$, are applied for time $t$}
\begin{equation} \label{t}
    t = \pi/\Omega_{R,n},
\end{equation}
where $\Omega_{R,n} = i \frac{a}{\omega_{1,n}}\left\langle\psi_1\left|(i \hbar \hat{p}_z )\right| \psi_n \right\rangle$ is the Rabi frequency \cite{mechoscillation}. Here $a$ is the vibration strength.  This prepares the neutron in an excited gravitational state $\ket{\psi_n}\ket{\uparrow_x}$. The resonance frequency between, for example, the first and fifth gravitational energy levels is $\omega_{1,5} = 5.15\times 10^{3}$ rad/s, while the Rabi frequency for a vibration strength of $a = 7 \; \mathrm{ms^{-2}}$ is $\Omega_{R,5} \approx 41$ rad/s.  This frequency corresponds to Rabi oscillation periods on the order of 25 milliseconds, which is achievable with mechanical systems~\cite{mechoscillation, natureexperiment}. 


In the third region, a constant magnetic field aligned along the z-axis is instantaneously switched on, preparing the neutrons in the state ${1/\sqrt{2}(c_{n}^{\uparrow}\ket{\psi_{n, \uparrow}}\ket{\uparrow_z} + c_{n}^{\downarrow}\ket{\psi_{n,\downarrow}}\ket{\downarrow_z})}$, where $c_{n}^{\uparrow(\downarrow)}$ are complex amplitudes in the expansion of the initial state $\ket{\psi_n}$ in the eigenbasis of the Hamiltonian in eq~\eqref{general}.  However, as shown in Appendix \ref{airy_a}, changes to gravitational eigenstates (due to  switching on the magnetic field) produce effects of order $\delta^2$ in the interferometric scheme considered here and can therefore be neglected.  The neutrons approximately remain in the $n$th gravitational energy level, i.e.~the total state reads ${1/\sqrt{2}(\ket{\psi_{n}}\ket{\uparrow_z} + \ket{\psi_{n}}\ket{\downarrow_z})}$.  Therefore, up to first order in $\delta$ we can make the approximation $\ket{\Psi_{n,s}} \approx \ket{\psi_{n}}\ket{s} $ .

\onecolumngrid

\twocolumngrid

The above prepared state evolves in the presence of the horizontal mirror at $z=0$ and a constant magnetic field along the z-axis  $\ket{\Psi\left(t \right)} = e^{-i\hat{H}t/\hbar}\frac{1}{\sqrt{2}}(\ket{\Psi_{n,\uparrow}} + \ket{\Psi_{n,\downarrow}})$, where $\hat H$ is given by  eq~\eqref{general}. This evolution causes the spin to precess but without exerting a force and thus without changing the gravitational state, producing a time-dependent relative phase $\theta$:
\begin{equation} \label{phase}
    \ket{\Psi\left(\theta \right)} = \frac{1}{\sqrt{2}}(\ket{\Psi_{n,\uparrow}} + e^{i\theta}\ket{\Psi_{n,\downarrow}}).
\end{equation}
The phase difference is directly proportional to the difference in energy between the neutron's spin up and spin down eigenstate $\theta = \frac{t}{\hbar}(E_{n, \uparrow}^{total} - E_{n, \downarrow}^{total})$ (and a global phase is ignored). As displayed in eq.~\eqref{energy_total}, the eigenenergies of the neutron {depend} on both relativistic corrections to their gravitational energy as well as the difference in the energy between spin up and spin down states of the neutrons. As a result, $\theta$ takes the form:
\begin{equation} \label{phase_diff}
    \theta = \frac{t}{\hbar}\left(\hbar\omega_0 +  \Delta E_r \right),
\end{equation}
{where} $\Delta E_r$ denotes the energy difference between {the} eigenstates due to relativistic corrections
\begin{equation} \label{rel_energies}
\begin{split}
   \Delta E_r &=  E_{n,\uparrow} - E_{n, \downarrow} = \frac{2}{3}\delta E_n.
\end{split}
\end{equation}

The relative phase expressed in eq.~\eqref{phase_diff} can be determined from the probability to measure the neutron's spin along the x-direction $\ket{+}= \frac{1}{\sqrt{2}}(\ket{\uparrow_z} + \ket{\downarrow_z})$;
and the probability reads
   $ p(\theta) = \Tr\left\{\ket{\Psi\left(\theta \right)}\bra{\Psi\left(\theta \right)}\ket{+}\bra{+}\right\} = \frac{1}{2}(1 + \cos(\theta))$.
The probability of measuring the state $\ket{+}$ has an explicit time dependence:
\begin{equation} \label{time}
    p(t) = \frac{1}{2}\left(1+ \cos\left(\frac{t}{\hbar}\left(\hbar\omega_0 +  \Delta E_r \right)\right)\right).
\end{equation}
As a result, experimental measurement of the phase $\theta$ with enough precision to measure the correction $\frac{t}{\hbar}\Delta E_r$ would allow us to determine whether mass-energy equivalence holds and the quantised spin energy contributes to the inertial and to the gravitational masses of the neutron. Thus, more generally, the experiment can also serve as a test of the EEP in the quantum regime~\cite{QEP,Orlando2016}. 

{We note that} $\hbar\omega_0 \gg \Delta E_r$, since $\Delta E_r \sim10^{-15}$ peV, while $\hbar\omega_0\approx 5\times10^{6}$ peV for a neutron in a $45$ T magnetic field (see section \ref{sec4}). As a result, the interference pattern which includes the correction from the mass-energy equivalence will be slightly shifted relative to the non-relativistic interference pattern.

Including the corrections to the eigenstates, the probability of measuring the state $\ket{+}$ is 
\begin{equation}
	\begin{split}
	p(t)=\frac{1}{2}\left(1+ A(t) \cos \left(\omega t\left(1+\frac{E_{n}}{3 m c^{2}}\right)\right)\right),
	\end{split}
\end{equation}

 As shown in Appendix~\ref{airy_a}, $A(t)$ to second order is $A(t) \approx 1-\left(\omega_{0} t\right)^{2}\left(\frac{E_{n}}{2 m c^{2}}\right)^{2}$. Therefore, the quantised spin energy contributions to the inertial and gravitational masses of the neutron can also be observed as a reduction in visibility at second order in $\delta$. However, the smallness of the relativistic correction delta ($\delta \sim 10^{-15}$) suggests this effect will not impact current interferometric experiments.

\section{Metrological enhancement in the estimation of the relativistic correction to the frequency}\label{sec:QFI}
Our proposal from the previous section for the detection of mass-energy effects in neutrons using {spin interferometry} is the most intuitive experimental procedure which is in principle capable of detecting these effects. However, a simple measurement of the phase between the spin states ignores information potentially present in the motional DOF. When considering more general measurements, the approximations in Section~\ref{sec5}, where we neglected the dynamics of the motional DOF, might not hold. Indeed, for similar settings, it has been shown that, if the dynamics of the motional DOF are taken into account, then higher measurement precision can be achieved \cite{PhysRevA98}. Using this approach, here we investigate whether higher sensitivities than measurements on solely the neutron's spin states (and thus simply a measurement of interferometric phase) can be achieved. 


We begin by formulating the task of detecting the effects of mass-energy equivalence in neutrons as a quantum parameter estimation problem. To simplify the analysis, rather than considering the dynamics of a gravitationally bound neutron, we consider a neutron that has an initial state described by the gravitationally bound eigenstates in eq.~\eqref{airy}, but with its dynamics described by the Hamiltonian of a freely falling particle in a gravitational potential. We expect this to be a good approximation at short times, as long as the wave packet has not fallen far from its initial height and thus the presence of the mirror can be neglected. We will support this expectation a posteriori for the initial states of interest, through a numerical calculation that includes the full Hamiltonian. Mathematically, we expect the approximation to hold for the dynamics of a gravitationally bound neutron in an eigenstate with negligible downward momentum for short times such that $t \ll \sqrt{\frac{2\bra{\psi_n}\hat{z}\ket{\psi_n}}{g}}$, in which case the free-fall dynamics of the neutron's CoM are negligble. The short time regime is of interest because it was found in Ref.~\cite{PhysRevA98} that the metrological enhancement from considering the motional DOF only exists at short times.

Under the above assumptions, we represent the relativistic correction as a perturbation to the neutron's non-relativistic Hamiltonian, by expanding ${\hat M = m\hat{I}+\hbar\omega_0\hat\sigma_z/2c^2}$ in the neutron's Hamiltonian ${\hat{H} = \hat{M}c^2+\frac{\hat{p}^2}{2\hat{M}} + \hat{M}g\hat{z}}$ to lowest order in $1/c^2$ and removing the constant offset $mc^2$:
\begin{equation} \label{HamiltonianQFI}
    \hat{H} \approx\frac{\hbar\omega_0}{2}\hat\sigma_z +\left(\frac{\hat{p}^2}{2m} +mg\hat{z} \right) + \delta\hat\sigma_z\left(-\frac{\hat{p}^2}{2m} + mg\hat{z} \right).
\end{equation}

The task is now formulated as the {measurement of the} value of $\delta$. As examined in section \ref{sec5}, the relative phase accumulated in a matter-wave interferometer between gravitationally bound spin up and spin down neutrons, $\phi = \frac{2t}{3\hbar}\delta E_n$ where $n$ is the gravitational energy level of the neutron.  Assuming $N$ uncorrelated particles are detected at the output of the interferometer, a population difference measurement on the spin states yields the following sensitivity:
\begin{equation} \label{backofh}
    \Delta \delta' = \frac{1}{\sqrt{N}}\left(\frac{2t}{3\hbar}E_n \right)^{-1}.
\end{equation}
{We stress that} eq.~\eqref{backofh} gives the sensitivity to which we can measure $\delta$ assuming a population difference measurement solely on the spin states of the system. However, this formula does not reveal the ultimate sensitivity to which $\delta$ can in principle be measured -- which is obtained from the quantum Cramer-Rao bound~\cite{Braunstein1994}:
\begin{equation} \label{Cramer}
    (\Delta \delta)^2 \geq \frac{1}{N F_Q},
\end{equation}
where $F_Q$ is the quantum Fisher information (QFI), which contains the full metrological information relevant for the {measurement of $\delta$. {For pure states, QFI is defined by}
 \begin{equation} \label{QF2}
    F_Q =4\left(\left\langle\mathrm{d}_{\delta} \psi(t) | \mathrm{d}_{\delta} \psi(t)\right\rangle-\left|\left\langle\mathrm{d}_{\delta} \psi(t) | \psi(t)\right\rangle\right|^{2}\right).
\end{equation}
In our case, $\ket{\psi(t)}=\exp\left(\tau(\hat{H}_0 + \delta\hat{H}' )\right)\ket{\psi(0)}$, $\tau := -\frac{it}{\hbar}$, and $\hat{H}' = \hat\sigma_z\left( -\frac{\hat{p}^2}{2m} + mg\hat{z}\right)$ denotes the perturbation due to relativistic effects.

Applying the Baker-Campbell-Hausdorff (BCH) formula, {up to a global phase}, we obtain: 
\begin{equation} \label{BCHexpansion}
e^{\tau(\hat{H}_0 + \delta \hat{H}')} \propto e^{\tau\hat{H}_0}e^{\tau\delta \hat{H}'}e^{\frac{-\tau^2}{2}[\hat{H}_0, \delta\hat{H}']} e^{\frac{\tau^{3}}{6}([\hat{H}_0,[\hat{H}_0, \delta\hat{H}']])}.
\end{equation}
Here we have neglected the term $e^{\frac{\tau^{3}}{6}(2[\delta\hat{H}' ,[\hat{H}_0, \delta\hat{H}']])}$, because $[\delta\hat{H}' ,[\hat{H}_0, \delta\hat{H}']] \propto i\hat{I}$ and therefore simply results in a global phase, which cancels when the expression is applied to eq.~\eqref{QF2} for the calculation of the QFI. Higher order terms in the BCH expansion commute and therefore eq.~\eqref{BCHexpansion} represents the expansion of $e^{\tau(\hat{H}_0 + \delta \hat{H}')}$ for all $\tau$. 
The commutation relations $\left[\hat{H}_0, \delta\hat{H}'\right] = -i{2\delta g\hbar}\hat\sigma_z\hat{p}$ and $[\hat{H}_0, [\hat{H}_0, \delta\hat{H}']] = 2mg^2\hbar^2 \hat{\sigma}_z \delta$  allow us to represent the time evolved state as
\begin{equation}\label{eq:fullstateQF}
\ket{\psi(t)} = e^{\tau\hat{H}_0 }e^{\tau \delta \hat{\sigma}_z \left( -\frac{\hat{p}^2}{2m} + mg\hat{z} + tg\hat{p} - mg^2 \frac{t^2}{3}\right)}\ket{\psi(0)}.
\end{equation}
While $\hat{p}$ does not commute with $\delta\hat{H}'$, their commutation is a complex constant. Therefore, we have been able to combine $e^{\tau\delta \hat{H}'}e^{\frac{-\tau^2}{2}[\hat{H}_0, \delta\hat{H}']}$ into one exponential while neglecting the commutation relation between the exponents, since this commutation relation will simply result in a global phase. In addition, we can combine the exponential $e^{\frac{\tau^{3}}{6}([\hat{H}_0,[\hat{H}_0, \delta\hat{H}']])}$ into this exponential because $\hat{\sigma}_z$ commutes with $\delta \hat{H}'$ and $\left[\hat{H}_0, \delta\hat{H}'\right]$.

Using eq~\eqref{eq:fullstateQF} in formula \eqref{QF2}, the QFI has a simple expression:
\begin{equation} \label{fullqf}
    F_Q = \frac{4t^2}{\hbar^2}\text{Var}\left(\hat{\sigma}_z \left( \frac{-\hat{p}^2}{2m} + mg\hat{z} + tg\hat{p} - m g^2 \frac{t^2}{3}\right) \right),
\end{equation}
{where $\text{Var}(\hat{A}) := \langle \hat{A}^2 \rangle - \langle \hat{A}\rangle^2 $}, and the expectation value is taken with respect to the initial state $\ket{\psi \left(0 \right)}$, such that $\langle \hat{A}\rangle = \bra{\psi \left(0 \right)} \hat{A} \ket{\psi \left(0 \right)} $. We now solve for the initial state $\ket{\psi \left(0\right)} = \frac{1}{\sqrt{2}}\ket{\psi_n}\left(\ket{\uparrow_z} + \ket{\downarrow_z} \right)$, where we have applied that up to first order in $\delta$, corrections to the gravitational eigenstates  of $\hat{H}_0$ due to the perturbation $\delta \hat{H}'$ are negligible (Appendix \ref{airy_a}). The initial state contains a superposition of spin states which results in $\langle \sigma_z^2 \rangle = 1$ and $\langle \sigma_z \rangle = 0$. These expectation values simplify the expression for the QFI in eq.~\eqref{fullqf} to be, 

\begin{equation} \label{finalQF}
\begin{split}
	F_Q &= \frac{4t^2}{\hbar^2} [\bra{\psi_n}  \left( \frac{-\hat{p}^2}{2m} + mg\hat{z} + tg\hat{p} \right)^2 \ket{\psi_n} \\
	&-\frac{2mg^2}{3}t^2 \bra{\psi_n}\left( \frac{-\hat{p}^2}{2m} + mg\hat{z}\right)  \ket{\psi_n} + \frac{m^2g^4}{9}t^4].
\end{split}
\end{equation}
Here, the expectation value is with respect to the Airy function eigenstates of $\hat{H}_0$. As shown in Appendix \ref{appA}, the Fourier transform of Airy functions produces momentum space wave functions which have an even probability amplitude. Hence, we have $\langle \hat{p}^3\rangle = 0$. Additionally, we use the numerically calculated result that $\langle \hat{z}\hat{p} + \hat{p}\hat{z}\rangle = -\frac{i\hbar}{2} + \frac{i\hbar}{2} = 0$ in eq.~\eqref{fullqf}, to obtain the following expression for the QFI:
\begin{equation} \label{QFexpanded}
\begin{split}
F_Q &= \frac{4t^2}{\hbar^2}\left[ \langle \left(\frac{\hat{p}^2}{2m}\right)^2  \rangle +  \langle\left(mg\hat{z}\right)^2\rangle \right.\\ 
    & \left. -\frac{g}{2}(\langle\hat{p}^2\hat{z} \rangle +  \langle\hat{z}\hat{p}^2 \rangle)  + t^2g^2\langle \hat{p}^2 \rangle \right.\\
   & \left. -\frac{2mg^2}{3}t^2 \left( \left\langle -\frac{\hat{p}^2}{2m} \right\rangle + mg \langle \hat{z} \rangle \right) + \frac{m^2g^4}{9}t^4  \right].
\end{split}
\end{equation}
The most useful quantity for comparing the ultimate precision predicted by the quantum Cramer-Rao bound and the precision obtainable with the simple spin interferometry is the ratio 
\begin{equation}
	\left(\frac{\Delta \delta'}{\Delta \delta}\right)_n = \frac{3}{2}\sqrt{\frac{F_Q\hbar^2 }{t^2E_n^2}}.
\end{equation}
This ratio is a dimensionless factor that describes the improvement in sensitivity that can be achieved by a measurement that saturates the quantum Cramer-Rao bound for neutrons in a particular gravitational eigenstate, compared to the corresponding measurement on neutrons in the same gravitational eigenstate with simple spin interferometry.  

If eq.~\eqref{backofh} provides the ultimate limit to which the effects of mass-energy equivalence can be measured, then the QFI should be quadratic in time at all time scales so that $(\Delta \delta'/\Delta \delta) \sim 1$. However, the QFI has several terms which do not change quadratically in time. This suggests that eq.~\eqref{backofh} may only provide the ultimate limit to which the effects of mass-energy equivalence can be measured on a particular timescale. Therefore, there may exist a timescale for which the effects of mass-energy equivalence can be measured with even higher precision. To investigate this, we define constants $\alpha_n$, $\beta_n$ and $\gamma_n$ to be coefficients for the terms in the QFI which do not change quadratically with time. Firstly, we define  $\alpha_n~:=~g^2 \langle \hat{p}^2 \rangle/E_n^2$ $\mathrm{s}^{-2}$ such that $t^2g^2\langle \hat{p}^2 \rangle \approx \alpha_n t^2 E_n^2$.  Therefore, $\alpha_n t^2$ is a dimensionless factor that scales quadratically with time. The subscript indicates the dependence of this constant on the gravitational eigenstate $\ket{\psi_n}$ which is used to compute the expectation value. Similarly, we define $\beta_n := \frac{2mg^2}{3 E_n^2}\left( \left\langle -\frac{\hat{p}^2}{2m} + mg \hat{z} \right\rangle \right)$ $\mathrm{s}^{-2}$, such that $\beta_n t^2$ is also a dimensionless factor that scales with time. Finally, we define $\gamma_n :=  \frac{m^2 g^4}{9 E_n^2}$ $\mathrm{s}^{-4}$.
 
  The numerically calculated sum of all other terms in eq.~\eqref{QFexpanded} is $4\frac{t^2}{\hbar^2} \left \langle \left(\frac{-\hat{p}^2}{2m} + mg\hat{z}\right)^2 \right \rangle \approx 1.9 \frac{t^2}{\hbar^2}E_n^2$. The total QFI is:
\begin{equation} \label{FQ}
F_Q \approx \frac{t^2}{\hbar^2}[1.9 E_n^2 + 4 t^2 E_n^2 (\alpha_n - \beta_n + t^2\gamma_n)].
\end{equation}
 Using eq.~\eqref{FQ} in the quantum Cramer-Rao bound, eq.~\eqref{Cramer}, gives the ultimate precision to which $\delta$ can be measured
\begin{equation} \label{highsensitivity}
(\Delta \delta)_n \geq \frac{1}{\sqrt{N} \sqrt{\frac{t^2}{\hbar^2}[1.9 E_n^2 + 4t^2 E_n^2 (\alpha_n  - \beta_n + t^2\gamma_n)]}}.
\end{equation}

In  {the} regime where $t \ll 1 \; \text{ms}$,  the first term of eq.~\eqref{FQ} dominates such that 
\begin{align}
F_Q^{ST} &\approx 4\frac{t^2}{\hbar^2}\left \langle \left(\frac{-\hat{p}^2}{2m} + mg\hat{z}\right)^2 \right \rangle \approx 1.9 \frac{t^2}{\hbar^2} E_n^2, \label{FQ_approx}
\end{align} 
and the term $tg \hat{p}$ in equation \eqref{fullqf} {can be neglected}. Here we see that at short times a more optimal measurement would give a factor 2 improvement over the sensitivity predicted by eq.~\eqref{backofh} such that $\left(\frac{\Delta \delta^{\prime}}{\Delta \delta}\right)_{n} \approx 2.1$. To verify that this approximation holds for bounded neutrons, we calculate the QFI from the quantum state obtained by numerically propagating the (unperturbed) ground state by projecting it into the ($\delta$-dependent) airy-function basis (see Fig.~\ref{fig:QFI_numeric}). We find that equation (\ref{FQ_approx}) holds up to $t \approx 0.5$ms despite neglecting the reflecting boundary. This is consistent with our original expectation regarding the approximation of the full dynamics with the free-fall one. Indeed, $0.5$ ms corresponds to a free-fall distance of the order of a $\mu \mathrm{m}$, while the height expectation value of the ground state is $\sim 9 \mu \mathrm{m}$ (and $\langle \hat{p}\rangle = 0$, see Appendix~\ref{appA}).

\begin{figure}[h] 
\begin{center}
\includegraphics[width =0.8\columnwidth]{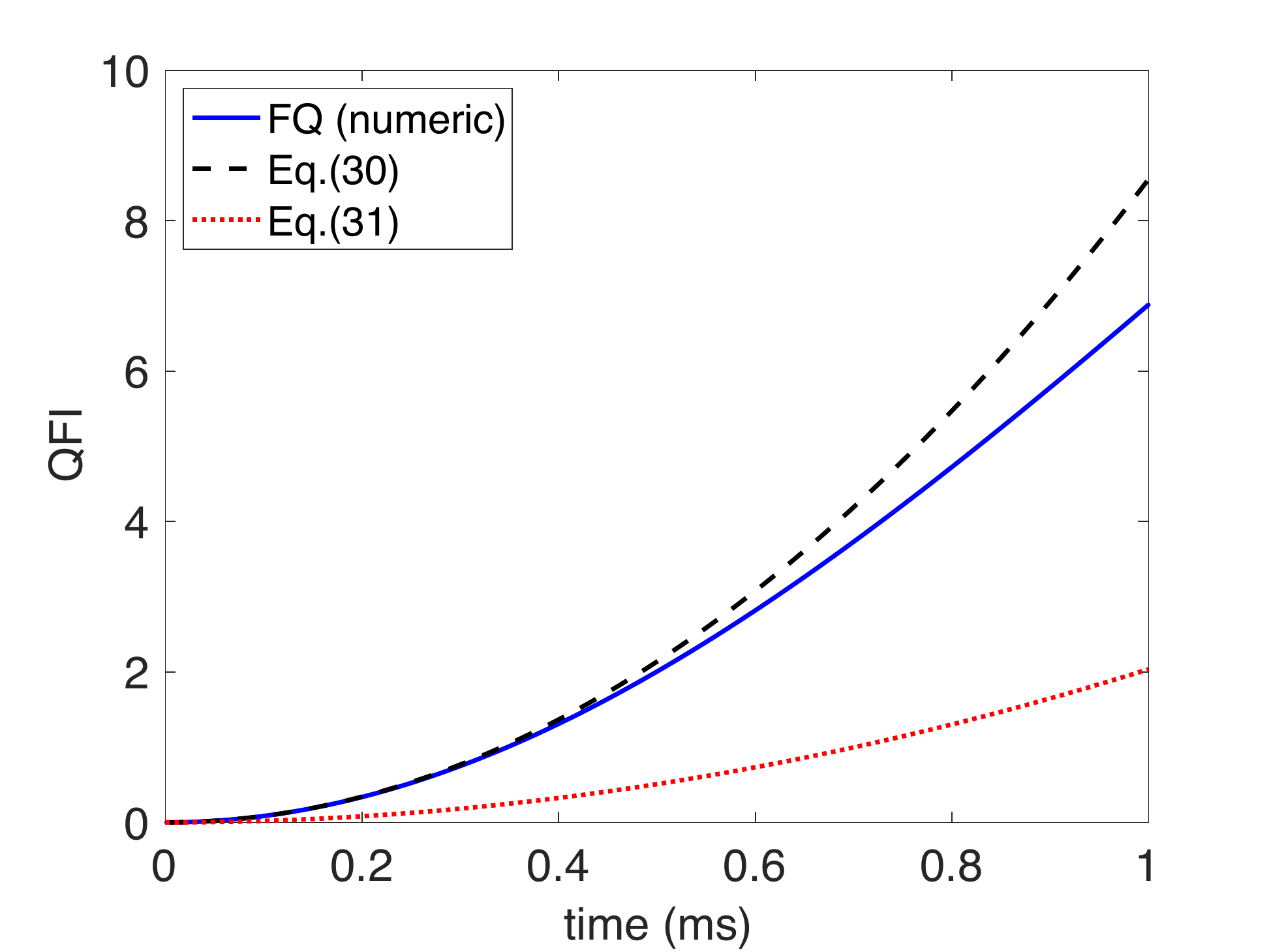}
\caption{\label{fig:QFI_numeric} The QFI calculated numerically (solid blue line) compared to equation (\ref{FQ_approx}) (dashed black line), and equation (\ref{scQFI}) (red dotted line). The initial state was chosen as the gravitational ground state. }
\end{center}
\end{figure}

In order to fully extract this sensitivity, we can adapt the general procedure presented in \cite{Macri:2016, Nolan:2017b, Haine:2018b}, which involves reversing the initial state-preparation dynamics and then making a  measurement that projects into the basis of which the initial state is an element. This can be achieved by switching off the magnetic field after time $t$, and making a measurement that can resolve both the $x$ component of the spin, and determines the motional energy. The additional information comes from the small probability of the particle being found in a different energy eigenstate due to the different rest masses resulting in a non-stationary state. It is important to note that a measurement of the external DOF that is precise enough to sufficiently resolve the motional energy is required to extract the optimal sensitivity. If only the spin component is measured, we recover the sensitivity from eq.~\eqref{backofh}.
 
 
 The recovery of equation \eqref{backofh} in the regime $t \sim 1 \; \text{ms}$ can be confirmed through a semi-classical calculation of the QFI which neglects the non-commutativity of position and momentum. As shown in section \ref{sec5}, unitary evolution of an initial superposition of spin states in a Mach-Zender interferometer simply produces a phase  $\ket{\Psi(t)} = \frac{1}{\sqrt{2}}(\ket{\Psi_{n,\uparrow}} + e^{i\frac{t}{\hbar}(\hbar\omega_0 + \frac{2}{3}\delta E_n)}\ket{\Psi_{n,\downarrow}})$. Neglecting the {commutator} between position and momentum, {and using} the above state in eq.~\eqref{QF2} gives
\begin{equation} \label{scQFI}
F_Q^{SC} = \frac{4t^2E_n^2}{9\hbar^2} ,
\end{equation}
which is fully consistent with eq.~\eqref{backofh}.

\section{Neutrons in Free-Fall}\label{sec_freefall}
The $t^6$ term in eq.~\eqref{FQ} indicates that enhanced sensitivity to the effects of mass-energy equivalence exists at long times for neutrons in free-fall\footnote{A similar $t^6$ scaling in the QFI was found independently in the context of measurement of acceleration~\cite{timesense}. }. Here, we investigate this term in the QFI to gain a complete understanding of the timescales for which higher sensitivities than that obtainable from a measurement on solely the neutron's spin states can be achieved. We begin with a freely falling neutron spin-polarised in the $z$-direction, before we initiate the interferometry sequence by using a magnetic field aligned in the $y$ direction to rotate the spin to be aligned in the $x$-direction. A strong magnetic field in the $z$ direction will now cause splitting in the energy of $|\uparrow_z\rangle$ and $|\downarrow_z\rangle$ components, which will contribute to the mass-energy of the system. We begin by using the stationary phase approximation to find an approximate expression for the phase, before considering the exact solution for a particular motional state. The classical action for a freely falling particle of mass $m$ initially at rest in a gravitational field $g$ is
\begin{align}
S_{cl} &= \frac{1}{3} m g^2 t^3 \, .
\end{align}
Under the stationary phase approximation, the state will then evolve to 
\begin{align}
|\Psi(t)\rangle &= \frac{1}{\sqrt{2}}\left(|\uparrow_z\rangle e^{-i \frac{\omega_0}{2} t}e^{i\phi_+}  + |\downarrow_z\rangle e^{i \frac{\omega_0}{2} t}e^{i\phi_-}\right)
\end{align}
where $\phi_{\pm} = \frac{1}{3} \frac{m\left(1\pm \delta\right) g^2 t^3}{\hbar}$ is the phase shift resulting from the motional contribution to the dynamics. Assuming the contribution from the magnetic field can be compensated for, the resulting phase difference is 
\begin{align} \label{phig}
\phi_g &= \frac{2\delta}{3} \frac{m g^2 t^3}{\hbar} \, .
\end{align}
Again, this phase can be detected by observing the sinusoidal phase dependence in the $x$ or $y$ component of the spin. Assuming $N$ neutrons are detected and the experiment is repeated $a$ times, we find the smallest value of $\delta$ with a population difference measurement on the spin states is
\begin{align}
\Delta \delta &= \frac{1}{\sqrt{aN}}\frac{3}{2}\frac{\hbar}{m_0 g^2 t^3}, \label{delta_freefall_sc}
\end{align}  
which has different scaling with the interrogation time $t$ compared to the bound neutron case. However, in writing equation (\ref{delta_freefall_sc}), we have neglected the fact that the motional wavepackets of the two spin-components will evolve slightly differently, due to the slight difference in their rest-mass. Specifically, denoting the motional degrees of freedom by $|\psi\rangle$, an initial state of the form  
\begin{align*}
|\Psi(0)\rangle &= \frac{1}{\sqrt{2}}\left(|\uparrow_z\rangle + |\downarrow_z\rangle\right)\otimes |\psi_0\rangle \label{psi_init}
\end{align*}
will evolve to
\begin{align}
|\Psi(t)\rangle &= \frac{1}{\sqrt{2}}\left(e^{-i \frac{\omega_0}{2} t}|\uparrow_z\rangle|\psi_\uparrow(t)\rangle  +  e^{i \frac{\omega_0}{2} t}|\downarrow_z\rangle|\psi_\downarrow(t)\rangle\right)
\end{align}
where $|\psi_{\uparrow, \downarrow}(t)\rangle = \exp\left(\frac{-i t}{\hbar}\left(\frac{\hat{p}^2}{2 m_{\uparrow, \downarrow}} + m_{\uparrow, \downarrow} g \hat{z}\right)\right)|\psi_0\rangle$. Using an initial state of the form
\begin{align}
|\psi_0\rangle &= \frac{1}{\sqrt{\sigma \sqrt{\pi}}}\int_{-\infty}^\infty e^{-\frac{z^2}{2\sigma^2}}|z\rangle dz \label{psi_motional}
\end{align}
and using the well known propagator for a particle in free-fall \cite{propagator}
\begin{align}
K(x, t, x^\prime, 0) &= \sqrt{\frac{m}{2\pi i \hbar t}}e^{\left[\frac{im t}{2\hbar}\left(\left(\frac{x - x^\prime}{t}\right)^2 - g(x+x^\prime) - \frac{1}{12}(gt)^2\right)\right]}
\end{align}
which with $ m_{\uparrow} = m(1 + \delta)$, or $m_{\downarrow} = m(1 - \delta)$ we use to propagate the sate for each internal mass energy, we find
\begin{align}
\langle \psi_\uparrow (t) | \psi_\downarrow (t)\rangle &= \mathcal{C} e^{i \phi_g}
\end{align}
where $\mathcal{C} = 1 - \mathcal{O}(\delta^2) \approx 1$ for the minuscule values of $\delta$ considered here, and $\phi_g$ is given in eq.~\eqref{phig}. The non-unity value of the overlap is due to the mass dependence in the dispersion relation. 


 Choosing our initial state of the form eq.~(\ref{psi_init}) yields $\langle \sigma_z^2 \rangle = 1$ and $\langle \sigma_z \rangle = 0$. We calculate the expression for the QFI in eq.~\eqref{fullqf} to be
\begin{equation} \label{finalQF}
	F_Q = \frac{4t^2}{\hbar^2}\langle \psi_0|\left( \frac{-\hat{p}^2}{2m} + mg\hat{z} + tg\hat{p} - m g^2 \frac{t^2}{3} \right)^2|\psi_0\rangle
\end{equation}
Here, the expectation values are calculated with respect to the initial motional state $|\psi_0\rangle$. For an initial state of the form \eqref{psi_motional}, this reduces to
\begin{equation}
F_Q = \left(\frac{2 m g t}{\hbar}\right)^2\left(\frac{\sigma^2}{2} + \frac{2}{3} \left(\frac{\hbar t}{\sigma m}\right)^2 + \frac{3}{16} \frac{\hbar^4}{\sigma^4 g^2 m^4} +\frac{g^2 t^4}{9} \right) \, .
\end{equation}
For large $t$, such that the free-fall distance ($L_g = \frac{1}{2} g t^2$) is large compared to both the width of the initial state ($L_{is} = \sigma$), and the spread in the width after time $t$ due to dispersion ($L_d \approx \frac{\hbar t}{\sigma m}$), and noting that the third term is $\frac{3}{16} \left(\frac{L_d}{L_g}\right)^2 L_d^2$, we can ignore all but the final term, and our expression reduces to
\begin{equation} \label{freefallQFI}
F_Q \approx \frac{4}{9} \frac{m^2 g^4 t^6}{\hbar^2} \, ,
\end{equation}
consistent with eq.~(\ref{delta_freefall_sc}). A similar result at long times has recently been independently derived in the context of measurement of acceleration~\cite{timesense}.

This analysis highlights that the $t^6$ term in the QFI  arises because of the time dilation-induced phase shift between the internal spin states. To see this, we note that under the stationary phase approximation (which we have shown holds at long interference times), the internal spin state of the freely falling neutron is
\begin{align}
|\Psi(t)\rangle &= \frac{1}{\sqrt{2}}\left(|\uparrow_z\rangle  + |\downarrow_z\rangle e^{i\omega_0t-i\frac{2}{3} \frac{\delta m g^{2} t^{3}}{\hbar}}\right).
\end{align}
A calculation of the QFI using eq.~\eqref{QF2} with this initial state reproduces eq.~\eqref{freefallQFI}. Therefore this term can be extracted by a measurement of the internal spin DOF. Our analysis shows that although the $t^6$ term in the QFI present in the previous section highlights an enhanced sensitivity to mass-energy equivalence at long times, the supplementary measurement of the motional DOF only provides non-negligible additional information in the regime of short times. The additional information comes from the quantised mass-energy induced coupling between the system's spin state and its motional energy state, which has the effect that at short times additional QFI is contained within the motional DOF, while at long times most of the QFI is simply contained within the phase.



\section{Conclusion}
We developed a theoretical framework to study the mass-energy effects in gravitationally bound neutrons. The consideration of the interaction of neutrons' spin with an external magnetic field provided a physical scenario in which the quantum effects of mass-energy equivalence become relevant. We found that while the quantised gravitational energy levels of the neutron are on the order of peV, in the presence of a 45 T magnetic field relativistic corrections cause shifts to these energy levels on the order of $10^{-15}$ peV.

We proposed a spin-interferometric experiment to detect the above effects in gravitationally bound neutrons. While challenging due to the required high sensitivity, the proposed experiment has the potential to {provide a new} test whether the EEP holds {for a system whose mass effectively becomes a quantum operator -- due to spin-dependent internal energy of the neutron in an external magnetic field}. We found through quantum parameter estimation, that it is in principle possible to design an optimised experiment with enhanced sensitivity to the relativistic correction to the precession frequency in the experimentally relevant regime of short interference times. The metrological enhancement only exists at short times and arises due to the mass-energy-induced coupling between the internal and external DOF of the neutron. To extract this metrological enhancement, a joint measurement of both the internal spin state and the external motional energy of the neutron should be performed. 

\section*{Note added:}
In the final stages of the preparation of our manuscript, we became aware of a parallel independent work that is related to section~\ref{sec:QFI} and section~\ref{sec_freefall} of our manuscript \cite{timesense}. 

\section*{Acknowledgements}
G.T.\ acknowledges support from a Cambridge Australia Allen STEM scholarship, jointly funded by Cambridge Australia Scholarships and the Cambridge Trust. F.C.\ acknowledges support through an Australian Research Council Discovery Early Career Researcher Award DE170100712. S.H. acknowledges support through an Australian Research Council Future Fellowship grant FT210100809. M.Z. acknowledges support through an Australian Research Council  Discovery Early Career Researcher Award DE180101443 and a Future Fellowship grant FT210100675. F.C.and M.Z. acknowledge support through ARC Centre of Excellence EQuS CE170100009. We acknowledge the traditional owners of the land on which the University of Queensland is situated, the Turrbal and Jagera people. We acknowledge the traditional owners of the land on which the Australian National University is situated, the Ngunnawal people.

\appendix
\section{Binomial expansion of the relativistic corrections to the gravitationally bound neutron's eigenenergies} \label{app2}
Here we derive a simple expression for the relativistic correction to a gravitationally bound neutron's eigenenergies in the presence of a constant external magnetic field. We start by considering the eigenergy of a gravitationally bound neutron with a shifted mass which accounts for mass-energy equivalence in a spin up neutron:
\begin{equation}
\phi\left(-\frac{\sqrt[3]{2}}{\sqrt[3]{m(1 + \delta)g^2\hbar^2}}E_{n,\uparrow}\right) = 0.
\end{equation}
Here, $\delta = \frac{\hbar \omega_0}{2mc^2}$ is the dimensionless relativistic correction. Expressing $E_{n,\uparrow}$ in terms of the zeros of the Airy function $\gamma_n$, we obtain:
\begin{equation}
	E_{n,\uparrow} = -\gamma_n\frac{\sqrt[3]{m(1 + \delta)g^2\hbar^2}}{\sqrt[3]{2}} .
\end{equation}
We can further simplify by the application of a binomial approximation for $ \delta \ll 1$: 

\begin{equation}
	E_{n,\uparrow} = -\gamma_n\frac{\sqrt[3]{mg^2\hbar^2}}{\sqrt[3]{2}}\left(1+ \frac{\delta}{3}\right) .
\end{equation}
Here, we recognize from eq. \eqref{energies} that $E_n =  -\gamma_n\frac{\sqrt[3]{mg^2\hbar^2}}{\sqrt[3]{2}}$ is the eigenenergy of a gravitationally bound neutron without the consideration of relativistic effects. This allows us to express the eigenenergies of a gravitationally bound neutron which account for mass-energy equivalence in terms of $E_n$ and $\delta$,
\begin{equation}
	E_{n,\uparrow} = E_n\left(1+ \frac{\delta}{3}\right).
\end{equation}
Repeating this derivation for a spin down neutron, we obtain an identical expression, but with an opposite sign for the relativistic correction:
\begin{equation}
	E_{n,\downarrow} = E_n\left(1- \frac{\delta}{3}\right).
\end{equation}

\section{Relativistic corrections to the gravitational eigenstates} \label{airy_a}
Here we show that corrections to the interferometric probability eq.~\eqref{time} due to the change of eigenstates are negligible at first order in $\delta$. We start by considering the effects of mass-energy equivalence as a perturbation to the gravitational eigenstates of the neutron. The Hamiltonian of the system $\hat{H}$, can  be expressed as the sum of the unperturbed Hamiltonian $\hat{H}_0 =\frac{\hbar\omega_0}{2} \hat\sigma_z+ \frac{\hat{p}^2}{2m} + mg\hat{z}$, and the perturbation due to mass energy equivalence $\hat{H}' = \hat\sigma_z\left( -\frac{\hat{p}^2}{2m} + mg\hat{z}\right)$.
\begin{equation}
	\hat{H} = \hat{H}_0 + \delta\hat{H}'.
\end{equation}
We first apply time-independent, non-degenerate perturbation theory \cite{LeBellac2006} to determine the corrections to the neutron's eigenvalues, showing they agree with our simple binomial expansion. 

In general, the first order correction to the eigenvalue $E_{n,\uparrow}^{(1)}$ is given by 
\begin{equation} \label{energy_corr1}
E_{n,\uparrow}^{(1)} = \bra\uparrow\bra{\psi_n}H'\ket{\psi_n}\ket\uparrow
\end{equation}
where $\ket{\psi_n}\ket{\!\!\uparrow}$ is the eigenstate of the unperturbed Hamiltonian, and in particular $\ket{\psi_n}$ is the eigenstate of $\frac{\hat{p}^2}{2m} + mg\hat{z}$. Recall that the virial theorem states that for a bound eigenstate of a  Hamiltonian which includes a potential $V(z)$ satisfying $z\, dV/dz = k V$, $k\in\mathbb{N}$, the following holds: $2 \langle \hat{p}^2/2m \rangle = k  \langle V(z) \rangle$. In our case  $V = mg\hat z$ and $k=1$ which allows us to obtain
\begin{equation} \label{energy_corr2}
E_{n,\uparrow}^{(1)} = \frac{1}{3}E_{n}^{(0)}, 
\end{equation}
where $E_{n}^{(0)}\equiv E_{n}$ is the energy without spin-dependent corrections. In other words, the total energy, up to $\mathcal{O}(\delta)$, reads $E_{n, \uparrow}= E_n^{(0)}+\delta E_{n,\uparrow}^{(1)} = E_n^{(0)}\left(1+\frac{\delta}{3}\right)$. 
The calculation for the opposite spin projection is fully analogous with only the sign difference: $E_{n, \downarrow} \approx E_n^{(0)}\left(1-\frac{\delta}{3}\right)$,  due to $\bra\downarrow\hat\sigma_z\ket\downarrow=-1$.

We now show that corrections to the eigenstates are negligible in the final interferometric probability $p= \Tr\left\{\ket{\Psi\left(\theta \right)}\bra{\Psi\left(\theta \right)}\ket{+}\bra{+}\right\}$, see  main text. Note that this probability can be  written in terms of the conditional state after the spin projection $p=\frac{1}{4}|| e^{-itH^\uparrow/\hbar}\ket{\psi_n} + e^{-itH^\downarrow/\hbar}\ket{\psi_n}||^2$ where $H^{\uparrow}:=\bra\uparrow H\ket\uparrow$, and analogously for $H^{\downarrow}$. We stress that $\ket{\psi_n}$ is the initial n-th gravitationally bound eigenstate and thus not an eigenstate of $H^{\uparrow (\downarrow)}$. The resulting $p= 1/2+  \bra{\psi_n}e^{itH^\uparrow/\hbar} e^{-itH^\downarrow/\hbar}\ket{\psi_n}/4+c.c$  can always be written as
\begin{equation}\label{probab_with_vis}
p(t)=  \frac{1}{2}\left( 1+ \left|\bra{\psi_n}e^{itH^\uparrow/\hbar} e^{-itH^\downarrow/\hbar}\ket{\psi_n}\right|\cos(\phi)\right)
\end{equation}
for some $\phi$.
Expanding the exponentials to second order we obtain
\begin{equation} \left| \langle e^{itH^\uparrow/\hbar} e^{-itH^\downarrow/\hbar} \rangle\right|  = 1-\frac{\Delta_n^2t^2}{2\hbar^2}
\end{equation}
where 
\begin{equation}
\Delta_n^2 = \langle (H^\uparrow-H^\downarrow -\delta \hat c_pt)^2 \rangle - \langle H^\uparrow-H^\downarrow -\delta \hat c_pt \rangle^2
\end{equation}
and the expectation values are taken in the state $\ket{\psi_n}$, and where
 $\hat c_p := mg\hat p $ comes from the non-commutativity of $H^\uparrow, H^\downarrow$. Due to the fact that $\langle p\rangle =0$ for Airy functions (see also Appendix C) we have
 \begin{equation}\label{approx_visib}
\Delta_n^2 = \langle (H^\uparrow-H^\downarrow)^2 \rangle - \langle H^\uparrow-H^\downarrow \rangle^2 +\mathcal{O}(\delta^2).
\end{equation}
Furthermore, $\langle (H^\uparrow-H^\downarrow)^2 \rangle - \langle H^\uparrow-H^\downarrow \rangle^2$ is itself of second order in $\delta$, specifically it reads $4\delta^2(\langle {H'}^2 \rangle - \langle{H'}\rangle^2)$. This shows that 
$p(t)\approx\frac{1}{2}\left( 1+ \cos(\phi)\right)$ and thus to lowest order in $\delta$ the relativistic effects indeed can at most give contributions to the interferometric phase.  At this order we have $\phi \approx \frac{t}{\hbar}\langle (H^\uparrow-H^\downarrow)\rangle$. Using again the virial theorem we obtain $\langle (H^\uparrow-H^\downarrow)\rangle = \hbar\omega +\frac{2}{3}\delta E_n$, which gives the phase  written in the main text.

The relation $\phi \approx \frac{t}{\hbar} \langle (H^\uparrow-H^\downarrow)\rangle$ can be directly obtained by expanding 
$p(t)\approx ||\left(1-i \frac{t}{2\hbar}(H^\uparrow +H^\downarrow)-\left( \frac{t}{2\hbar}\right)^2({H^\uparrow}^2+{H^\downarrow}^2)\right)\ket{\psi_n}||^2$. To second order this yields $p(t)\approx 1-\left(\frac{t}{2\hbar}\right)^2\langle(H^\uparrow +H^\downarrow)^2\rangle$. Comparing this to eq.~\eqref{probab_with_vis}: $\frac{1}{2}+\frac{1}{2}(1-t^2\Delta_n^2/2\hbar^2)\cos(\phi)$ we see that $\cos(\phi)\approx 1-t^2/2\hbar^2\langle(H^\uparrow +H^\downarrow)\rangle^2$ which indeed gives $\phi= \frac{t}{\hbar}\langle(H^\uparrow +H^\downarrow)\rangle$

\section{Proof of expectation value of momentum for real wavefunctions} \label{appA}
Here we prove the result for the airy function eigenstates $\psi_n(z)$, $\langle \hat{p}^n \rangle = 0$, for all odd positive integers $n \in 1,3,5,...$ We begin the proof by taking the Fourier transform of the position space wavefunctions: 
\[
\begin{split}
\phi_{n}(v) &= \sqrt{\frac{m}{2 \pi \hbar}} \int_{0}^{\infty} \psi_{n}(z) \exp \left(-\mathrm{i} \frac{m v z}{\hbar}\right) \mathrm{d} z \\
&= \sqrt{\frac{m}{2 \pi \hbar}} \int_{0}^{\infty} \psi_{n}(z) \left[ \cos\left(\frac{mvz}{\hbar}\right) -i\sin\left(\frac{mvz}{\hbar}\right) \right]\mathrm{d} z \\
\end{split}
\]
As a result, if we calculate the complex conjugate of the momentum space wavefunction, it is simple to calculate the probability amplitude in momentum space:
\[
\begin{split}
|\phi_{n}(v)|^2 &= \phi^*_{n}(v)\phi_{n}(v)\\
&= \frac{m}{2\pi\hbar} \left[\int_{0}^{\infty} \psi_{n}(z)\cos\left(\frac{mvz}{\hbar}\right)\mathrm{d} z \right]^2 \\
&+ \frac{m}{2\pi\hbar} \left[\int_{0}^{\infty} \psi_{n}(z)\sin\left(\frac{mvz}{\hbar}\right)\mathrm{d} z \right]^2 .
\end{split}
\]
This result relies on the assumption that $\psi_n(z)$ is real, which is true for the airy function stationary states we use in this paper. Here it is easy to see that $|\phi_{n}(-v)|^2 = |\phi_{n}(v)|^2$, thus completing the proof that the probability amplitude in momentum space is even, allowing us to conclude that $\langle \hat{p}^n \rangle = 0$, for all odd positive integers $n$.

\bibliography{bibliography} 
\bibliographystyle{ieeetr}

\end{document}